\documentstyle[11pt]{article}
\begin{document}
\pagestyle{plain}
\huge
\title{\bf The rocket equations for decays of elementary particles}
\large
\author{Miroslav Pardy\\[7mm]
Department of Physical Electronics \\
and\\
Laboratory of Plasma physics\\[5mm]
Masaryk University \\
Kotl\'{a}\v{r}sk\'{a} 2, 611 37 Brno, Czech Republic\\
e-mail:pamir@physics.muni.cz}
\date{\today}
\maketitle
\vspace{5mm}

\begin{abstract}

The decay of elementary particles is  described
nonrelativistically and  using the method of the special theory of
relativity. Then the Tsiolkovskii rocket equation is applied to the
one photon decay of the excited nucleus of the M\"ossbauer effect.
The formation time of photons during decay is supposed nonzero. The Me\v s\v
cerskii equation is possible to identify with the bremsstrahlung equation.
All decays described in the ``Review of Particle physics properties''
can be investigated from the viewpoint of the rocket equations

\end{abstract}

\vspace{3mm}


\vspace{3mm}

\section{Introduction}

Many elementary particles are not stable and can decay to the
different components as it is shown in 
``Review of Particle physics properties'' (Alvarez-Gaum\'e, 2004).
To be pedagogically clear we first describe the classical decay of
particles. We follow the text of Landau et al. (Landau et al. 1965),
where the decay of elementary particles is described
nonrelativistically and  using the methods 
of the special theory of relativity (Landau et al. 1962).
The quantum field theory describes decays of elementary
particles only using the asymptotic states. 
Decay is stimulated (under the influence of the external
forces), or, spontaneous (under the influence of internal forces,
or hidden external forces). 
The most simple description of decay is in the system where
the original particle was in the state of rest. With regard to the
validity of the momentum conservation law, the particles have the same
magnitude of the momentum, however they moves in the opposite
direction.
 
If the decay is photonic, then we suppose that  
the Tsiolkovskii rocket equation can be applied to the
one photon decay of the excited nucleus. First, we apply it 
to the the M\"ossbauer process.
The formation time of photons during decay is supposed nonzero as an 
analog to the Einstein idea (Einstein, 1926 ). 
If electron is accelerated then it radiates. 
We show the Me\v s\v
cerskii equation is possible to identify with the bremsstrahlung
equation where electron is in the permanent decay.
Many other decays can be investigated from the view point of the Me\v s\v
cerskii equations. 
We think that the decays described in the 
``Review of Particle physics properties'' (Alvarez-Gaum\'e, 2004)
can be investigated  from the viewpoint of the rocket equations. The
article is the original trial to unify the cosmonautic physics with
particle physics.   

\section{The decay of the nonrelativistic particles}

Let as consider a decay of some  particle to two particles 1 and
2. The most simple is to consider the decay in the rest system of the
original particle. The conservation law of momentum is valid during 
the decay and it means that particles are emitted with the same momenta in
the opposite directions.  
Let us denote the magnitude of the  momentum of every
particle as $p_{0}$. Then the energy conservation laws reads:
$$E_{int} = E_{1int} + \frac{p_{0}}{2m_{1}} + E_{2int} +
\frac{p_{0}}{2m_{2}}, \eqno(1)$$
where $m_{1}, m_{2}$ are particle masses and $ E_{1int},  E_{2int}$
are their internal energies and $ E_{int}$ is the internal energy of
the original decaying particle. If $\varepsilon$ is the energy of the
decay, then evidently

$$\varepsilon  = E_{int} -  E_{1int} -  E_{2int} \eqno(2)$$
and this energy must be positive in order to be the decay possible.

It follows from eq. (2) and (1) that 

$$\varepsilon = \frac{p_{0}^{2}}{2}\left(\frac{1}{m_{1}} + 
\frac{1}{m_{2}}\right)= \frac{p_{0}^{2}}{2m},  \eqno(3)$$
where $m$ is the reduced mass of both particles. The velocities of
every particle are $v_{10} = p_{0}/m_{1}$ and  $v_{20} = p_{0}/m_{2}$. 

Now, let us consider the so called the laboratory system L where the
the original particle moves with the velocity of the magnitude
$V$. Let us consider one of the particle and denote its velocity as
${\bf v}$ in the L-system and ${\bf v_{0}}$ in the CMS. Then,

$${\bf v} = {\bf V} + {\bf v_{0}}, \quad {\rm or}, \quad 
{\bf v}-  {\bf V} = {\bf v_{0}}  \eqno(4)$$
and 

$$v^{2} + V^{2} - 2vV\cos\theta = v_{0}^{2}, \eqno(5)$$
where $\theta$ is the angle of escape of the particle with regard to
the direction of the velocity ${\bf V}$.

If the angle of escape of he particle in the CMS system is
$\theta_{0}$, then there is an elementary relation between $\theta$ and 
$\theta_{0}$. Or, (Landau et al., 1965):

$$\tan\theta = \frac{v_{0}\sin\theta_{0}}{v_{0}\cos\theta_{0} + V}. 
\eqno(6)$$

If we solve this equation in order to get $\cos\theta_{0}$, the we
obtain after some elementary operations (Landau et al., 1965):

$$\cos\theta_{0} = -\frac{V}{v_{0}}\sin^{2}\theta \pm \cos\theta
\sqrt{1 - \frac{v^{2}}{v_{0}^{2}}\sin^{2}\theta}. \eqno(7)$$

If the decay angles are $\theta_{1}$ and $\theta_{2}$ in the L-system, 
then there is the following relation between  $\theta_{1}$ and
$\theta_{2}$ (Landau et al., 1965):

$$\frac{m_{2}}{m_{1}}\sin^{2}\theta_{2} + 
\frac{m_{1}}{m_{2}}\sin^{2}\theta_{1}  - 2\sin\theta_{1}\sin\theta_{2}
\cos(\theta_{1} + \theta_{2}) = \frac{2\varepsilon}{(m_{1} +
  m_{2})V^{2}}\sin^{2}(\theta_{1} + \theta_{2}). \eqno(8)$$

\section{The decay of the relativistic particles}

Let us consider, following Landau et al. (1962) the relativistic 
spontaneous decay of a
body of mass $M$ into two particles with mass $m_{1}$ and $m_{2}$. 
Let us consider the decay in the system where the body is at rest.  
The law of the conservation of the relativistic energy is 

$$M = E_{10} +  E_{20}, \eqno(9)$$
where $E_{10}$ and $E_{20}$ are the relativistic energies of the
emerging particles. Here the velocity of light is $c = 1$. Since
$E_{10} > m_{1}$ and $E_{20} > m_{2}$, the equality (9) can be
satisfied only when $M > m_{1} + m_{2}$, i.e. a particle can
disintegrate  spontaneously into two parts when the sum of their
masses is less than the mass of the original particle. On the other
hand, when $M < m_{1} + m_{2}$, the original particle is stable and
does not decay spontaneously. To cause the decay in this case, we must
to supply to the original particle some energy from outside. The  
applied energy must be at least the binding energy $m_{1} + m_{2} -
M$.

The momentum must be also conserved in the decay process. Since the
initial momentum of the original particle was zero, the sum of the
momenta of the emerging particles must be also zero, or ${\bf p_{10}}
+  {\bf p_{20}} = 0$. Consequently ${p_{10}}^{2} =  {p_{20}}^{2}$, or 

$${E_{10}}^{2} - m_{1}^{2} =   {E_{20}}^{2} - m_{2}^{2}. \eqno(10)$$

The to equations (9) and (10) uniquely determine the energies of the
emerging particles:

$$E_{10} = \frac{M^{2} + m_{1}^{2} -  m_{2}^{2}}{2M}, \quad 
E_{20} = \frac{M^{2} -  m_{1}^{2} +  m_{2}^{2}}{2M}. \eqno(11)$$

Now, let us suppose that  the original particle is moving with the
velocity $V$ and dissociates ``in flight'' into two particles with
mass m.
Let us determine the relation between angle of emergence of
these particles and their energies. Let be $E_{0}$ the energy of one
particle in the CM-system. In other words the energy is given by the
formula (11). If $\theta$ is the angle of the emergence in the
L-system, then using the relativistic transformation formulas, we get
(Landau et al., 1962):

$$E_{0} = \frac{E - Vp\cos\theta}{\sqrt{1 - V^{2}}},\eqno(12)$$
from which follows 

$$\cos\theta = \frac{E - E_{0}\sqrt{1 - V^{2}}}{V\sqrt{E^{2} - m^{2}}}.
\eqno(13)$$

For determination of $E$ from $\cos\theta$ we then get the quadratic
equation.

$$E^{2}(1 - V^{2}\cos^{2}\theta)- 2EE_{0}\sqrt{1 - V^{2}} + E_{0}^{2}
(1 - V^{2}) + V^{2}m^{2}\cos^{2}\theta = 0, \eqno(14)$$
which has one positive root (if the velocity $v_{0}$ of the decay
particle in the CM-system  satisfies $v_{0} > V$), or two positive
roots if $v_{0} < V$. The more detailed discussion of this case is
described in Landau et al. (1962) textbook. 

Now, let us consider the situation where the particle with mass $M$ in
rest decays into the system of the three particles $m_{1}, m_{2},
m_{3}$. Let us answer the question, what is the maximal energy of
particle $m_{1}$ of the three-body system. 
  
The solution of this problem is as follows (Landau et al. 1962). The
particle  $m_{1}$ has its maximum energy if the system of the other
two particles $m_{2}$,  $m_{3}$, is the least possible mass which is the
sum  $m_{2}$ +  $m_{3}$ and corresponds to the case where the rest
particles moves together with the same velocity. So, we can reduce the
problem to the decay of a body into two parts, and so we get using 
eq. (11) the following expression (Landau et al. 1962):

$$E_{1max} = \frac{M^{2} + m_{1}^{2} -  (m_{2} + m_{3})^{2}}{2M}.
\eqno(15)$$
 
\section{The Tsiolkovskii  rocket equation for decay}

There are some experiments which prove the particle properties of a
photons. On the other hand photon moves as a wave with the wave length
$\lambda$ and moves with velocity $c$ in vacuum. The particle-wave 
synthesis can be expressed by the relation $p = \hbar\omega/c$ where 
$p$ is a momentum of the photon and $\omega$ is its frequency. The
mass of photon follows from the Einstein energetic relation $E =
mc^2$ as $m = \hbar\omega/c^2$.

However, when an excited nucleus $N^*$, (for instance of the M\"ossbauer
effect) decays according to the
equation $N^* \rightarrow N + \gamma$, then during the very short
interval photon is created and it has no particle and wave
properties. The problem of  the process of emission of photon was
firstly formulated by Einstein (1926). He tried to decide if photon is
created at one moment, or  as a wave during some formation time. This
Einstein work has continuation in the  the blackbody article where Einstein
supposes only the quanta of electromagnetic energy, later denoted
by photons, but he does not consider the formation time of photons.   
(Einstein, 1917). From the Einstein period, 
photon is the most mysterious particle in universe.

We here suppose that there is a formation  time of a photon 
and it can be evidently
postulated as $T = \lambda/c$ as the minimal time for the formation of
the wave length $\lambda$. During the formation time the nucleus $N$ is
accelerated according to the equation $v = at$, where $a$ is
acceleration and the final velocity $v_f$ is $v_f = a\lambda/c$. The
velocity of the nucleus during the formation time of photon must be described by
the rocket equation by Tsiolkovskii (1962):

$$v = c\ln\left(\frac{m_0}{m(t)}\right), \eqno(16)$$
where $c$ is the velocity of light and it is the velocity of the light
as the medium, $m_0$ is the initial mass of the nucleus and $m(t)$ is
the variable mass during the formation of a photon. The nucleus is micro-rocket during the formation time of photon. 

In the formation time $T = \lambda/c$ we can write

$$m = m_{0} - m_0\alpha t = m_0(1 -\alpha t) \approx m_0 e^{-\alpha t};
\quad 0 < t < \lambda/c .\eqno(17)$$

Then, from equation (16) and (17) it follows that 

$$v = c\alpha t. \eqno(18)$$

We have for the acceleration $dv/dt = a = c\alpha = v_{f}/T $, where 

$$v_{f} = \frac{\hbar\omega}{m_{0}c}. \eqno(19)$$

Then it easily follows that 

$$\alpha = \frac{1}{2\pi}\frac{\hbar\omega^{2}}{m_{0}c^{2}}. \eqno(20)$$
and then

$$m(t) = m_{0}e^{-\frac{1}{2\pi}\frac{\hbar\omega^{2}}{m_{0}c^{2}}t};
\quad 0\leq t \leq \lambda/c. \eqno(21)$$

In case that the mass of crystal in the  M\"ossbauer experiment 
 is $M_{0}$ and the time dependence of
mass of crystal is $M(t)$ and the recoil is transmitted to the crystal
as a whole, then instead of eq. (21) we write:

$$M(t) = M_{0}e^{-\frac{1}{2\pi}\frac{\hbar\omega^{2}}{M_{0}c^{2}}t};
\quad 0\leq t \leq \lambda/c .
\eqno(22)$$

So we have seen that the combination of the elementary relativity
equations with the Tsiolkovskii rocket equation, the decay of nucleus
is described during the formation time of photon. It also means that
the M\"ossbauer effect can be completed by the information on the
microscopical description of the decay of nucleus. In case that the
mass of the decaying object is the whole  crystal as it is supposed in
case of the low temperature M\"ossbauer situation, then, the mass
change of the crystal is practically zero. The quantum mechanical
probability that the crystal survive at the zero temperature at the
same state after emission of the photon with momentum $P$ is 

$$Probability \quad  = \quad \exp\left\{-\frac{P^{2}}{\hbar^{2}}
\langle Z^{2}_{M} \rangle \right\}, \eqno(23)$$ 
where $\langle Z^{2}_{M} \rangle$ is the average shift of the
emitting nucleus  from the equilibrium position (Feynman, 1972). 
The analogous expression is valid for the finite-temperature 
M\"ossbauer effect.

\section{The Me{\v s}{\v c}erskii rocket equation}

Let us consider the nucleus with mass $m$ moving with the velocity
${\bf v}$ with regard to the inertial system $S$ at time $t$ and the
force $d{\bf F}$ acts during the time $dt$ on the nucleus and during
this infinitesimal time the mass $dm_{1}$ is emitted with he velocity
${\bf u_{1}}$ and the mass $dm_{2}$ is emitted with he velocity
${\bf u_{2}}$. The equation of the infinitesimal motion of this system
is as follows:

$$d{\bf p} = d(m{\bf v}) = m d{\bf v} +  {\bf v} dm = m d{\bf v} +
{\bf v}(dm_{1} + dm_{2})\eqno(24)$$ 

At the same time, the following equation is correct:

$$  d(m{\bf v}) = {\bf F}dt + {\bf u_{1}}dm_{1} + {\bf u_{2}}dm_{2},
\eqno(25)$$     
which is the Newton law of action and reaction.

The equation (25) can be evidently written in the form 

$$\frac{d{\bf p}}{dt} = {\bf F} + {\bf u_{1}}\frac{dm_{1}}{dt} + 
{\bf u_{2}}\frac{dm_{2}}{dt}. \eqno(26)$$ 

Then, we have from eqs. (24) and (26) the so called 
Me{\v s}{\v c}erskii rocket equation (Me{\v s}{\v c}erskii, 1962,
Kosmodemiakskii, 1955, 1966):

$$m\frac{d{\bf v}}{dt}  = {\bf F} + ({\bf u_{1}} - {\bf v})\frac{dm_{1}}{dt} + 
({\bf u_{2}} - {\bf v})\frac{dm_{2}}{dt}. \eqno(27)$$ 

If we consider the emission of many masses $dm_{1},dm_{2}, ...dm_{N}$,
then we get instead of of equation (27), the following equation 

$$m\frac{d{\bf v}}{dt}  = {\bf F} + \sum_{i = 1}^{N}({\bf u_{i}} - 
{\bf v})\frac{dm_{i}}{dt} .\eqno(28)$$ 

In case that we consider only emission of one mass component and the
zero force acting on the accelerated nucleus, then
there is no mass term $dm_{2}$  an the motion is one-dimesional with
the corresponding Tsiolkovskii equation:

$$m\frac{d{ v}}{dt}  = { v_{r}}\frac{dm}{dt}\eqno(29)$$ 
where $v_{r}$ is the relative velocity of he emitted mass with the
regard to the nucleus. 

The corresponding solution can be easily find as follows:

$$v = v_{0} + v_{r}\ln\frac{m_{0}}{m(t)},\eqno(30)$$
where $m_{0}$ is the initial mass of the nucleus and $m(t)$ is its
instantaneous mass at time $t$.  

The Tsiolkovskii equation in the gravitational field follows from
eq. (29) evidently in the form 

$$m\frac{d{ v}}{dt}  = -mg + { v_{r}}\frac{dm}{dt},\eqno(31)$$ 
with the solution 

$$v = v_{0} - gt + v_{r}\ln\frac{m_{0}}{m(t)}.\eqno(32)$$
        
There are many nuclei which can be in the excited states and the
decays, cannot be necessarily one-photon decay but multi-photon decay.
$N^{*} \rightarrow \underbrace{\gamma + \gamma +    ....  \gamma}_{n}.$

In this case it is necessary to generalize the dependence of mass of
nucleus on time. We can  suppose that the decay starts 
at the same moment. Let us remark that the whole situation can be still
generalized to the case with massive photon and to the multi-photon
decay with the massive photons. Massive photons are physically 
meaningful as it is supposed in the author articles (Pardy, 2002, 2004).

\section{Decay of accelerated electron and rocket equations}

When electron (or every charged particle) is accelerated, then it
radiates the electromagnetic energy. The nonrelativistic equation of
motion involving the acceleration is as follows (Landau et al., 1988):

$$m\dot{\bf v} = {\bf F} + \frac{2e^{2}}{3c^{3}}\ddot{\bf v}.\eqno(33)$$

If we postulate the compatibility of the Me\v s\v cerskii equation with
the above equation, then, in the 1-dimensional case with $c>> v$, we
get after comparison of the these equations:

$$\frac{2e^{2}}{3c^{3}}\ddot{ v} = - c\frac{d\delta m}{dt}.\eqno(34)$$
from which follows the following dependence of emitted mass of the
decaying electron (being accelerated):

$$\delta m =  - \frac{2e^{2}}{3c^{4}}a.\eqno(35)$$

The interpretation of the result is that accelerated electron (or the
charged particle) is during  the acceleration in the permanent excited
state, or, 

$$e^{*} \rightarrow e + \gamma \eqno(36)$$ 
and the mass of the accelerated electron is renormalized according 
 the equation

$$m \to m + \delta m; \quad \delta m =  - \frac{2e^{2}}{3c^{4}}a.\eqno(37)$$ 
for all time of acceleration. It means that electron moving in the
magnetic field is in the permanent excited sate which permanently
decays.  The mass of decay is obtained by the absorptive mechanism
which was still not described in the classical textbooks on
electromagnetism. The quantum version of the absorptive mechanism is
also unknown. We think that the decay of the excited electron in the
magnetic field is not the same as the decay of the high-energy excited
electron investigated in the high-energy laboratories such as CERN.

\section{Decay of $K^{+}$  meson}

The decay of $K^{+}$-meson ($\theta$-decay) is radiative with emission of the
$\gamma$-photon, or nonradiative where the gamma-photon is not
present. So the nonradiative decay is (We do not consider the
$\tau$-decay expressed as $K^{+} \to \pi^{+} + \pi^{+} + \pi^{-}$,
which is the substantial ingredient of the CPT theorem, which was not
still investigated from the viewpoint of the rocket equations):

$$K^{+}(p) \to \pi^{+}(q_{1}) + \pi^{0}(q_{2})\eqno(38)$$
and the radiative decay is 
$$K^{+}(p) \to \pi^{+}(q_{1}) + \pi^{0}(q_{2}) + \gamma(k)\eqno(39)$$
 
The matrix element of the radiative decay of $K^{+}$-meson is
(Efrosinin, 2006):

$$\langle\pi^{0},\pi^{+},\gamma|S|K^{+}\rangle =
ie(2\pi)^{4}\delta^{4} (q_{1} + q_{2} + k - p)\varepsilon^{\mu}M_{\mu}(k) 
\eqno(40)$$
where

$$M_{\mu}(k) =  M_{\mu}^{photon\;emission\;from\;external\;line}(k)
\quad +\quad 
M_{\mu}^{direct\;vector\;contribution}(k) \quad + $$

$$M_{\mu}^{direct\;axial\;contribution}(k).\eqno(41)$$

Efrosinin(2006) show by calculation that the dominant contribution to
the K-meson decay cones from the bremsstrahlung term. 

However, because we suppose that the decay cam have the different form
which are not involved in the Feynman diagrams, we postulate the
possibility the following modification  of the K-meson decay:

$$K^{+}(p) \to \left(\pi^{+}(q_{1}) + \pi^{0}(q_{2})\right)
 + \gamma(k)\eqno(42)$$
and it means that this decay can be described by the rocket equations
from which follows the exponential decline of mass of the system 
$\left(\pi^{+}(q_{1}) + \pi^{0}(q_{2})\right)$. Let us still remark
 that the system $\left(\pi^{+}(q_{1}) + \pi^{0}(q_{2})\right)$ as
 a bound system can be described by the quark formalism.

\section{The alpha, electron, positron and  neutrino decays}

If the decay component is only gamma photons, then the decay is
massless because the rest mass of photon is zero. There are many
reactions which combine different types of decays (Rohlf, 1994). 
Let us write some examples, which can be also described by the 
quark formalism. 

$${^{12}C} \rightarrow {^{8}Be} + \alpha \eqno(43)$$ 
is massive decay because the $\alpha$-particle is massive
one. Similarly 
$${^{12}C} \rightarrow {^{8}Be} + \alpha , \quad 
{^{8}Be} \rightarrow \alpha + \alpha, \quad 
{^{210}Po} \rightarrow  {^{206}Po} + \alpha \eqno(44)$$
are massive decay. 

The combined decays are the following ones:

$${^{8}Be} \rightarrow  {^{8}Be} + e^{+} + \nu_{e} \eqno(45)$$
and 

$${^{15}O} \rightarrow  {^{15}N} + e^{+} + \nu_{e} \eqno(46)$$
are massive decay in combination 
with massless decay because we suppose that neutrino is massless
particle.

At present time it is not establish experimentally the process of decay
of Berylium and Oxygen. There is a possibility of
the following equations of the decays:

$${^{60}Co} \rightarrow  \left({^{60}Co} +   e^{-}\right) + \nu_{e} \eqno(47)$$
and 

$${^{15}O} \rightarrow  \left({^{15}N} +   e^{+}\right)  + \nu_{e} ,
\eqno(48)$$
which can be interpreted in such a way that the first step is the
  emission of neutrino and then the emission of positron. The
  experimental verification of such process can be performed with the
  inserting this process in the very strong magnetic field and then the
  process will be realized according to Me{\v s}{\v c}erskii equation as
  follows:

$$m\frac{d{\bf v}}{dt}  = \frac{q}{c}({\bf v}\times {\bf H}) + 
\sum_{i = 1}^{N}({\bf u_{i}} - {\bf v})\frac{dm_{i}}{dt}, \eqno(49)$$ 
where the external force of the equation (11) was replaced by the
Lorentz force for the motion of a charged particle in an
electromagnetic field.  To our knowledge, such experiment was not
still performed. This experiment can also be applied to the decay of
$\mu^{+}$-meson, because the decay equation is as follows
 
$$\mu^{+}\quad \rightarrow \quad e^{-} + \quad \nu_{e} + 
\quad \bar\nu_{e} .\eqno(50)$$ 

Similarly, we can consider the decay of excited electron according to
the equation $e^{*}\rightarrow e + \gamma,$ 
which is of course massless  decay. To our knowledge such decay
was not confirmed in the particle laboratories.

The decay of neutron is a such that in reality neutron and proton have
a rich substructure of quarks and gluons so that the $\beta$-decay
diagrams are oversimplified. However, at the quark level, the process

$$n \quad \rightarrow \quad p + e^{-} + \bar\nu_{e} \eqno(51)$$ 
can be considered as 

$$d \quad \rightarrow \quad u + e^{-} + \bar \nu_{e} \eqno(52)$$ 
i.e. d-quark changes  via quark-weak coupling to u-quark, an electron
and antineutrino. The decay (51) and (52) can evidently be considered
from the viewpoint of the rocket equations.

At he same time the decay of the Higgs particle can be considered 
from the viewpoint of the rocket equations. 
The same is valid for the broken quark confinement.
In the astrophysics, the nonsymmetrical supernova explosion from the
viewpoint of the rocket equation was not investigated till the
present time.

\section{Discussion}

We considered the decay of elementary particles nonrelativistically 
and  using the method of the special theory of
relativity. The asymptotic states of the quantum field theory was not 
considered. Then, the Tsiolkovskii rocket equation was applied to the
one photon decay of the excited nuclei of the M\"ossbauer effect.
The formation time of photons during decay was supposed to be
nonzero. The Me\v s\v cerskii equation was identified with 
the bremsstrahlung equation for electron in order to get that electron
is in the permanent decay during acceleration for instance when moving
in the electromagnetic field. Our theory concerns all 
decays described in the ``Review of Particle physics properties''
(Alvarez-Gaum\'e, 2004) and every decay can be 
investigated with regard to the rocket equations. The rocket equations
were still not used in the microphysics and in the
nanophysics. However, they are not prohibited.  
 
 From the viewpoint of the rocket equation  the mass of nucleus 
in the M\"ossbauer effect is exponentially formated . Similarly the
 possible bound state of elementary particle is formated exponentially.
The question of application of this fact to some 
calculations in QCD is open.  In QFT we use only
asymptotic states, which means that the description of reality is
incomplete. The theory of electroweak interaction must evidently work
with the formation time of particle mass and formation time of
photon. The formation time of photon was considered by Einstein (1926). 
The formation time of mass and  formation time of photon is reality. 
There is no quantum process with the zero formation time. 

How to confirm our theory by experiment? Careful observation of a single
electron in an atom trap over a period of several months  by Gabrielse
at al. has resulted in the best measurement yet of the electron's magnetic
moment and an improved value for alpha, the fine structure constant.
Why not use this method in the modified form for case where a charged 
particle radiates in the magnetic field, or in case of 
the M\"ossbauer process, or in case of decays of elementary particles?
 
Gerald Gabrielse and his students Brian Odom and David
Hanneke at Harvard combined
electric and magnetic forces to keep the electron in its circular
"cyclotron" orbit.  In addition to this planar motion, the electron
wobbles up and down in the vertical direction, in the direction of
the magnetic field (D'Urso et al., 2005).

The measured value of g can also be used to address the issue of
hypothetical electron constituents.  Such subcomponents could be 
no lighter than 130 GeV.  On the basis
of this experiment one can also place a corresponding limit on the
size of the electron; it must be no larger than $10^{-18}$ m across.
(Odom, et al. 2006).

We believe that the processes described in the above text 
and other processes of particle and nuclear physics 
can be experimentally confirmed by the modified Gabrielse et al. methods.
  
\vspace{15mm}

{\bf References}

\vspace{15mm}

\noindent 
Alvarez-Gaum\'e, L. et al. (2004). Review of particle physics,
 {\it  Phys. Lett}. {\bf 592}, No. 1-4.\\[2mm]  
 D'Urso, B. Van Handel, R., Odom, B.  and Gabrielse, G.(2005).
Single-Particle Self-excited Oscillator,  {\it Phys. Rev. Lett}. {\bf
  94}, 113002.\\[2mm] 
Efrosinin, V. P. (2006).  About radiative kaon decay $K^{+} \to
  \pi^{+}\pi^{0}\gamma$, hep-ph/0606216.\\[2mm] 
Einstein, A. (1917). Zur quantentheorie der Strahlung, 
{\it Physikalische Zeitschrift}, {\bf 18}, 121.\\[2mm] 
Einstein, A. (1926). Vorschlag zu einem die Natur des
 elementaren Strahlungs-Emissionsprocesses betrefenden Experiment, 
 {\it Naturwiss}, {\bf 14}, 300 - 301.\\[2mm] 
Feynman, R. (1972). {\it Statistical mechanics}, (W.A. Benjamin,
  Inc., Reading, Mass.).\\[2mm]
Kosmodemianskii, A.  A.  (1955). {\it The course of theoretical
  mechanics} I, (Moscow), (in Russian)., ibid. (1966). {\it The course of 
theoretical  mechanics} II, (Moscow), (in Russian).\\[2mm]
Landau, L. D. and Lifshitz, E. M. (1962). {\it The Classical
Theory of Fields}, 2nd ed. (Pergamon Press, Oxford).\\[2mm] 
Landau, L. D. and Lifshitz, E. M. (1965). {\it Mechanics}, 
2nd ed. (Nauka, Moscow). (in Russian).\\[2mm]
Me{\v s}{\v c}erskii, I. V.  (1962). {\it Works on
 mechanics of the  variable mass},  (Moscow). (in Russian).\\[2mm]
Odom, B., Hanneke D., D'Urso, B. and Gabrielse, G. 
New Measurement of the Electron Magnetic Moment Using a One-Electron 
Quantum Cyclotron,  {\it Phys. Rev. Lett.} {\bf 97}, 030801 (2006).\\[2mm] 
Pardy, M. (2002). \v Cerenkov effect with massive photons: 
{\it Int. Journal of Theoretical Physics} {\bf 41}, No. 5, 887.\\[2mm]
Pardy, M. (2004). Massive photons and the Volkov solution: 
{\it Int. Journal o f Theoretical Physics} {\bf 43}, No. 1, 127.\\[2mm]
Rohlf, J. W. (1994). {\it Modern Physics from $\alpha$ 
to $Z^{0}$}, (John Wiley \& Sons Inc. New York).\\[2mm] 
Tsiolkovskii, K. E. (1962).  {\it Collective works}, (Moscow).(in Russian). 

\end{document}